\documentstyle[12pt]{article}
\begin{document}
\begin{center}
{\LARGE\bf PHYSICS OF HEAVY QUARKS FROM LATTICE QCD} \\
\end{center}
\vspace{10mm}
\noindent
{\Large Apoorva Patel} \\
{\large CTS and SERC, Indian Institute of Science, Bangalore-560012} \\
\vspace{5mm}

\begin{abstract}
In the last few years, lattice QCD has made a dramatic progress in
understanding the physics of hadrons containing heavy quarks, from
the first principle. This review summarises the major achievements.
\end{abstract}
\vspace{10mm}

\centerline{\bf I. INTRODUCTION}

Heavy quarks play a dominant role in the understanding of the weak
interactions as well as of what may lie beyond the standard model.
Since quarks are always bound into hadrons, and we do not understand
the strong interactions rigorously, the results of these phenomena
are often expressed in terms of non-perturbative parameters (also
called matrix elements) that reflect our knowledge/ignorance of the
strong interaction effects. Lattice QCD offers the best route to a
non-perturbative determination of these parameters.

The field of lattice gauge theories has nowadays achieved a maturity
level where statistical errors are beaten down enough to expose
systematic effects. This control over systematic effects has helped
reliable extraction of matrix elements that allow us to test the
Standard Model. In fact the results of these numerical experiments
are now acknowledged by the Particle Data Group \cite{PDG96}.
What I present below are excerpts from results of many lattice
QCD calculations; more details can be found in the proceedings of
the annual lattice field theory conferences \cite{LAT95,LAT96}.

Lattice QCD simulations of hadrons containing heavy quarks provide
a testing ground for many new ideas. The heavy hadrons are smaller
than the light hadrons, and hence less susceptible to effects of
finite lattice volume and dynamical quarks. (Typical lattice size
in simulations nowadays is $1.5-2$ fm.) Their nonrelativistic
nature allows an easy comparison of the results with those from
phenomenological quark potential models. Moreover, the results
do not suffer from the uncertainties of chiral extrapolations.

Three different formalisms for heavy quarks have been used in
lattice QCD to get at the properties of heavy hadrons: (1) static,
(2) non-relativistic (NRQCD), (3) conventional.
\begin{eqnarray}
L_1 &=& \overline{Q} v_\mu D^\mu Q    \\
L_2 &=& \overline{Q} [ D^0 - {1 \over 2M}{\vec D}^2 + {c_1 \over 2M}
        \vec\sigma \cdot \vec B + {\cal O}({1 \over M^2}) ] Q \\
L_3 &=& \overline{q} (D\!\!\!\!/ + M) q
\end{eqnarray}
Here $Q$ refers to $2-$component heavy quark fields, defined so as to
remove the static mass $M$ from the Lagrangian (otherwise $M$ would
appear as an additive constant). The field $q$ contains both particle
and anti-particle components, and $M$ cannot be removed in that case.
The coefficient $c_1$ has to be non-perturbatively determined by a
matching between NRQCD and conventional formalisms.

In the static case, the $4-$velocity of the heavy quark is a constant
of motion, and the calculations are often carried out in the rest frame.
The formalism is meaningful when the hadron has only one heavy quark,
and the results provide the limiting behaviour as $M\rightarrow\infty$.
The NRQCD formalism is organised in powers of $M^{-1}$ for hadrons having
only one heavy quark, and in powers of $v \sim p/M \sim \alpha_s(p)$
for hadrons having more heavy quarks \cite{NRQCDexp}.

The lattice momentum cutoff in typical simulations nowadays is $\pi/a
\sim 5-10$ GeV. The NRQCD approach is useful for $Ma > 1$. With the
inclusion of ${\cal O}(M^{-1})$ terms, $L_2$ works well for $b-$quarks
and can be extrapolated to the case of $c-$quark. The conventional
approach requires $Ma < 1$. It works reasonably for $c-$quarks, and
can be extrapolated to heavier masses.

A good consistency check on the methods is that the NRQCD and the
conventional results agree in the region between $m_b$ and $m_c$.
This comparison requires various $M-$dependent renormalisations to
justify the extrapolations in the quark mass, but at the end we have
continuous functional behaviour for the quantities measured.

There are many technical details involved in converting lattice results
to numbers useful for phenomenology. The techniques are sophisticated,
and the caveats should be kept in mind, instead of blindly using the
numbers in phenomenological analyses. I only mention these without
elaboration:
\begin{itemize}
\item Both the action and the operators are improved, using higher
order terms in $a$ and $M^{-1}$ to cut down discretisation errors.
The creation and annihilation operators for the asymptotic states
are smeared to increase overlap with the desired wavefunctions.
\item The lattice perturbation theory is not well-behaved in terms of
the bare coupling. Rather one needs to express the results in terms of
the renormalised coupling that is extracted from the simulation itself.
A poor man's choice is ``tadpole-improved'' perturbation theory. With
a non-perturbative cutoff, the lattice results are unambiguous, but
renormalon ambiguities appear when converting them to a continuum scheme
such as $\overline{MS}$.
\item The symmetries of the lattice theory are only a subset of the
continuum ones. This makes renormalisation and mixing of the lattice
operators more complicated than in the continuum. In principle, all the
renormalisation constants should be determined through non-perturbative
definitions. In practice, it turns out that finite and logarithmic
renormalisations (e.g. dim$\le 4$ operators such as quark bilinears) can
be well estimated using the renormalised perturbation theory. Power law
divergences (e.g. dim$>4$ operators mixing with lower dimension operators),
however, have to be eliminated using non-perturbative subtractions based
on Ward identities.
\item A majority of lattice results, and certainly the ones where all the
errors are well under control, are obtained in the quenched approximation
ignoring the effects of dynamical quarks. This approximation violates
the unitarity of the theory and unphysical singularities appear in the
chiral limit. Simulation results indicate that these violations, however,
are not serious for typical simulations carried out with the light quark
about as heavy as the strange quark. For certain quantities, results
including two or four flavours of dynamical quarks have become available.
These are combined with the quenched (i.e. $n_f=0$) values to obtain
physically relevant numbers (i.e. $n_f=3$).
\item Specific scaling forms are used to extrapolate the results to the
infinite volume, continuum and chiral limits. Algorithms for simulation
of the path integral and for calculation of the quark propagator in a
background gluon field keep on getting revised. Parametrisation of the
highly correlated lattice data needs careful fitting procedures.
\end{itemize}

\bigskip
\centerline{\bf II. PROPERTIES OF HEAVY-HEAVY MESONS}

Spectroscopy of heavy quarkonia (Upsilon, Charmonium and $B_c$) is the
simplest test of lattice simulations. There are two bare parameters: the
coupling and the heavy quark mass. Effects of light quarks appear only
through vacuum polarisation loops. NRQCD simulations directly extract
the level splittings, and the lattice cutoff is determined (in GeV) by
assigning one of the splittings its physical value. The splittings are
largely insensitive to the precise value of $M$, which keeps systematic
errors under control. A remarkable feature of the results is that many of
them have been obtained on rather coarse lattices using improved actions.

Fig.1 shows how well the lowest spin-averaged states of the Upsilon
spectrum can be reproduced \cite{Shige}. A closer look reveals that the
effects of light dynamical quarks shift the numbers closer to their
experimental values. The Upsilon hyperfine splittings, shown in Fig.2,
are more sensitive to the details of the lattice action: quenching tends
to underestimate them, and accurate values are needed for $M$ and $c_1$
(which undergoes substantial renormalisation).

Results for Charmonium and $B_c$ (not yet discovered) spectra are not as
impressive. The main difficulty is that the quarks are more relativistic
and higher order terms in the NRQCD expansion are needed to keep the
systematic errors under control. The results are still phenomenologically
useful and can be used to tune quark models. An example of Coulomb gauge
wavefunctions for Charmonium states is shown in Fig.3 \cite{Lepage}.

Calculations of heavy-light hadron masses suffer from the additional
uncertainty of extrapolating the light quark mass to the chiral limit.
Results are again best expressed as level splittings. Reasonable values
are found for $B_s-B$, $\Lambda_b-B$ and $\Sigma_b-\Lambda_b$ splittings.
Hyperfine splittings (e.g. $B^*-B$ and $\Sigma_b^*-\Sigma_b$), however,
are typically underestimated and require more care.

First results for annihilation decays of quarkonia have recently become
available \cite{annihil1}. Annihilation decay rates, $\Gamma(Q\overline{Q}
\rightarrow light~hadrons, \gamma\gamma, l^+l^-)$, can be expressed as a
power series in $M^{-1}$, with each term factorised into a short distance
perturbative coefficient and a non-perturbative hadronic matrix element
of $4-$fermion operators \cite{annihil2}. In the vacuum saturation
approximation (VSA), the ``colour singlet'' combinations can be written
in terms of quarkonium wave functions at the origin and their derivatives.
The ``colour octet'' contributions measure the $Q\overline{Q}g$ component
in the $P-$wave states. The numerical results are in reasonable accord
with the VSA and the experimental data.

\bigskip
\centerline{\bf III. THE QCD RUNNING COUPLING}

To verify that QCD is indeed the theory of the strong interactions, it
must be demonstrated that perturbative and non-perturbative determinations
of the coupling $\alpha_s$ agree. The former is typically extracted from
experimental results of high energy processes. The latter can be evaluated
on the lattice from low energy hadronic properties. To this end, several
non-perturbative definitions of $\alpha_s$ have been employed (note that
$\alpha_{\overline{MS}}$ does not exist outside perturbation theory):
\begin{eqnarray}
\alpha_{q\overline{q}} (q={1 \over r}) &=& C_F^{-1} r^2 F(r)
       {\hskip 95pt}\stackrel{q\rightarrow0}{\longrightarrow}~ \infty \\
\alpha_{SF} (q={1 \over L}) &\propto& [ {\partial\over\partial\eta}
                                      \ln Z(L,\eta) ]_{\eta=0}
       {\hskip 61pt}\stackrel{q\rightarrow0}{\longrightarrow}~ 0 \\
\alpha_{TP} (q={1 \over L}) &\propto& {\langle P_x(0)P_x(L) \rangle
                                 \over \langle P_z(0)P_z(L) \rangle}
       {\hskip 80pt}\stackrel{q\rightarrow0}{\longrightarrow}~ {\rm constant} \\
\alpha_P (q={3.4 \over a}) &=& -{3 \over 4\pi}
                                {\ln \langle W_{1\times1} \rangle
                                \over (1-(1.19+0.07n_f)\alpha_P)}
       \stackrel{q\rightarrow0}{\longrightarrow}~ \infty
\end{eqnarray}
The infrared behaviour of these definitions differ, as indicated above,
but they should all come together towards the ultraviolet end. Strictly
speaking, with a non-perturbative regulator, scaling is exact only on
the critical surface. Away from $g=0$, scaling holds only upto a certain
tolerance level, and different definitions of $\alpha_s$ help quantify
this tolerance level. A comparison of $g^2_{SF}$ and $g^2_{TP}$ for the
pure gauge $SU(2)$ theory, shown in Fig.4 \cite{Weisz}, gives us faith 
to use perturbative scaling in the actual region of simulations.

Having demonstrated scaling, the lattice results can be converted into
a value for $\alpha_{\overline{MS}}(M_Z)$. The common procedure is to
use the spectrum results for quarkonia to determine $\alpha_P$ at a
specific value of the cutoff, and then use perturbation theory to
convert it into $\alpha_{\overline{MS}}$. Given that $\alpha_s={\cal O}
(0.1)$, a precision of ${\cal O}(1\%)$ requires $2-$loop perturbative
matching. Determinations of $\alpha_s$ from different spectral quantities
need not agree with each-other, when the number of dynamical quarks is
unphysical. It is reassuring to see, as illustrated in Fig.5 \cite{NRQCD},
that they indeed converge when extrapolated to $n_f=3$. Results obtained
by various groups and their extrapolations to $n_f=3$ are displayed in
Fig.6 \cite{Shige}; they typically lie at the lower edge of the band
of values from non-lattice determinations.

There is yet another step in converting lattice numbers to a physical
$\alpha_{\overline{MS}}(M_Z)$. The dynamical quarks included in lattice
simulations are much heavier than the physical $u$ and $d$ quarks, and
do not correctly reproduce the effects of pion loops. Chiral symmetry
breaking in QCD implies that extrapolation of spectral quantities to the
chiral limit be linear in the quark mass \cite{GrinRoth}, in contrast to
quadratic dependence seen in weak coupling perturbation theory. The final
result is \cite{Shige}
\begin{equation}
\alpha_{\overline{MS}}(M_Z) ~=~ 0.117(1)(1)(2) ~~,
\end{equation}
where the first error combines statistical and discretisation
uncertainties, the second is due to $m_{dyn}$, and the third comes
from perturbative conversions.

\bigskip
\centerline{\bf IV. PROPERTIES OF HEAVY-LIGHT PSEUDOSCALARS}

The pseudoscalar decay constant is a phenomenologically important
parameter. Though its calculation on the lattice is straightforward,
the number has remained hard to pin down due to its substantial mass
dependence---Fig.7 \cite{MILC} shows the extent of ${\cal O}(M^{-1})$
corrections to the static limit. Exploiting the freedom that the
various ${\cal O}(M^{-1})$ contributions can be selectively extracted
from lattice simulations, the high sensitivity of $f_{Qq}$ to quark
mass has been traced down to the large heavy quark kinetic energy
\cite{Collins}. Fig.8 \cite{MILC} shows the extrapolation of $f_B$
values to the continuum, and also indicates that dynamical quarks
would increase $f_B$ somewhat. Overall it is safe to quote
\cite{MILC,JLQCD}
\begin{equation}
f_B ~=~ 180(20)~{\rm MeV}~~({\rm with}~f_\pi=132~{\rm MeV}) ~~.
\end{equation}
The ratios of decay constants are more stable as several correlated
systematic errors cancel, and the results are \cite{MILC,JLQCD}:
\begin{equation}
f_B/f_D ~\approx~ 0.85 ~~,~~ f_{B_s}/f_B ~\approx~ 1.10 ~~,~~
                             f_{D_s}/f_D ~\approx~ 1.10 ~~,
\end{equation}
all with errors of about $5\%$.

Lattice calculations have also estimated several bound state parameters
often used in $B-$physics phenomenology. In absence of any asymptotic
quark states in QCD, these are defined through behaviour of lattice
quark propagators in the Landau gauge. To quote \cite{Gimenez}:
the $B-$meson binding energy $\overline{\Lambda}=180(30)$ MeV, the 
$\overline{MS}$ running quark mass $m_b(m_b)=4.15(20)$ GeV, the heavy
quark kinetic energy in $B-$meson $\lambda_1=0.1(1)$ GeV$^2$.

The $B-$parameter characterising $B-\overline{B}$ mixing is another
useful quantity. In the effective theory below $m_b$, it has no
anomalous dimension at $1-$loop. It can be easily seen, at least at
strong coupling, that the deviation of $B_B$ from one is ${\cal O}
(M^{-1},N_c^{-2})$.
Indeed different lattice calculations confirm that $B_B(m_b)=0.90(5)$
\cite{Flynn}. Dynamical quarks do not seem to have substantial effect
on $B_B$. Increasing the light quark mass increases $B_B$ slightly,
so that $B_{bs} > B_{bd}$. Conversion of the lattice $B_B$ to the
RG-invariant value used in continuum analyses is rather uncertain:
${\hat B}_B=1.3(2)$.

\bigskip
\centerline{\bf V. FORM FACTORS OF HEAVY-LIGHT HADRONS}

The spin-flavour symmetries of heavy quarks reduce form factors of
heavy-light hadrons to Isgur-Wise functions \cite{Neubert}. Matrix
elements of the charge operators (i.e. ${\vec p}_i = {\vec p}_f$)
determine the form factor normalisations, and provide scaling rules
for their $M-$dependence. In the heavy quark effective theory, the
charge operators act in the rest frame, and hence constrain the form
factors at $q^2=q^2_{\rm max}$. On the other hand, the charge operators
correspond to $q^2=0$ in the lightcone frame, and thus the lightcone
sum rules constrain the form factors at maximum recoil. The two
complementary constraints can be merged together in pole dominance
ans\"atze for the form factors, where the crucial property is the order
of the pole. Lattice calculations can be used to verify these ans\"atze.
In practice, it is easier to measure form factors near $q^2_{\rm max}$
in experiments, and near $q^2=0$ on the lattice. The pole dominance
ans\"atze are often conveniently adopted to extend them to other $q^2$.

Lattice calculations can observe how the symmetry relations amongst
the form factors get modified as the quark masses are reduced. For
the semileptonic pseudoscalar-to-pseudoscalar decay, the pole relation
between $f_+$ and $f_0$ is well satisfied, even when the the heavy
quark mass is ${\cal O}(m_c)$, as exemplified in Fig.9 \cite{LANL}.
The ${\cal O}(M^{-1})$ corrections to the relations amongst the form
factors appearing in the semileptonic pseudoscalar-to-vector decay are
larger, in particular the relation between $V$ and $A_1$ works poorly
even at $m_b$. The corrections seem to affect mostly the normalisations,
however, and not the $q^2-$dependence. The form factors occuring in
the radiative decay $B \rightarrow K^*\gamma$, and heavy-to-light
semileptonic decays $B \rightarrow \rho l \nu$ and $B \rightarrow \pi
l \nu$, can be related using heavy quark and light flavour $SU(3)$
symmetries. These relations seem to work well, but the $q^2-$dependence
of the form factors is not yet fully sorted out in lattice calculations
\cite{Flynn,Soni}.

The Isgur-Wise functions have been extracted from lattice form factors
evaluated at sufficiently heavy quark mass. Examples for meson and baryon
cases are shown in Fig.10 \cite{LANL} and Fig.11 \cite{UKQCD} respectively.
The meson Isgur-Wise function, $\xi(v_i \cdot v_f)$, shows an increase as
the light quark mass is reduced, and the slope at zero recoil is $\xi'(1)
\approx -1$ in the chiral limit. First results for the baryon Isgur-Wise
function \cite{UKQCD} indicate that it falls off more steeply than the
meson one, and its slope decreases by about a factor of two between $m_s$
and the chiral limit.

In strong coupling lattice QCD, $\xi(v_i \cdot v_f)$ has a double pole
form, with the singularity at $v_i \cdot v_f = -1$ \cite{PCVC}. This
result can be rephrased as PCVC: the divergence of the partially conserved
vector current is saturated by the scalar meson pole near $M_i+M_f$ (the
scalar pole may arise from the combination of a set of states spread over
an energy range much smaller than $M$). This ansatz fits both the lattice
and experimental (Fig.12 \cite{CLEOII} shows the latest results) data as
well as any other fit. It is worth exploring if there is any limit of QCD,
quite likely it would involve $N_c\rightarrow\infty$, where this ansatz
becomes exact.

\bigskip
\centerline{\bf VI. CONSTRAINING THE STANDARD MODEL}

Lattice results can help in precision tests of the standard model, by
pinning down the elements of the Cabbibo-Kobayashi-Masakwa quark mixing
matrix. Of particular interest is the unitarity triangle involving the
third generation, which determines $CP-$violation. The present constraints
can best be expressed in terms of the Wolfenstein parameters $\rho$ and
$\eta$, and they are displayed in Fig.13. Particularly useful here are
the lattice inputs for $f_B$, $B_B$ and $B_K$ \cite{Sharpe}. The bounds
suggest that, within the standard model, $CP-$violation is close to
maximal (i.e. the phase $\delta\sim\pi/2$). The standard model will be
constrained further by results leading to $|\epsilon'/\epsilon|$.

Lattice $f_B$, $B_B$ values also determine, in terms of $|V_{ts}|$, the
extent of $B_s-\overline{B_s}$ oscillations. These oscillations may be
observable at LEP, SLC and HERA-B \cite{Soni}.

\bigskip\noindent{\bf Acknowledgements:}
I thank the organisers for an informative and enjoyable workshop. I am
grateful to the Physics Department, University of Wuppertal for its
hospitality during the completion of this review.

\end{document}